\documentclass[fleqn,twoside]{article}
\usepackage{espcrc2}

\usepackage{epsf}

\newcommand{\be}{\begin{equation}}
\newcommand{\ee}{\end{equation}}

\title{Numerical results from large $N$ reduced QCD$_2$}

\author{J. Kiskis\address[UCD]{Department of Physics, University of
California, Davis, CA 95616}
        ,
        R. Narayanan\address{Department of Physics, Florida International
University, Miami, FL 33199}\thanks{Speaker}
,
and
        H. Neuberger\address{Department of Physics, Rutgers University,
Piscataway, NJ 08855}\thanks{The research of H.N. was supported in part
by DOE grant DE-FG02-01ER41165.}
}

\begin{document}

\begin{abstract}
Some results in QCD$_2$ at large N are presented using the reduced model
on the lattice. Overlap fermions are used to compute meson propagators.
\vspace{1pc}
\end{abstract}

\maketitle

\begin{figure}
\epsfxsize = 0.45\textwidth
\centerline{\epsfbox{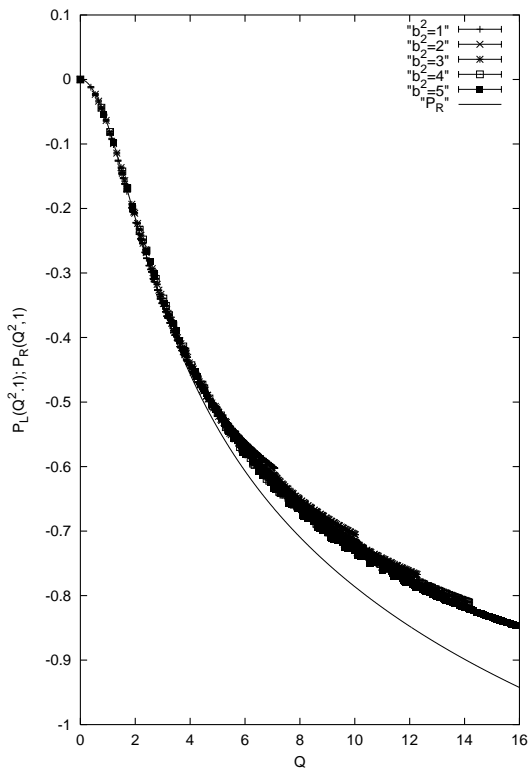}}
\caption{
Plot of $P_L(Q^2,1)$ at $N_c=41$ for $b^2=1,2,3,4,5$.
}
\label{pseudog1}
\end{figure}

\section{Introduction}

This follows the talk given by H. Neuberger~\cite{talk1} and summarizes
the numerical results obtained in QCD$_2$ at large N using the quenched
reduced model and overlap fermions~\cite{paper}.

The reduced model has a single site, and there are $d$ SU(N) matrices
$ U_\mu,\mu=1,\cdots,d$. The gauge action is given by
\be
S_g = - \beta \sum_{\mu > \nu} Tr C_{\mu\nu} C^\dagger_{\mu\nu};\ \ \ \
C_{\mu\nu} = [U_\mu, U_\nu],
\label{gauge}
\ee
where
$\beta = {1\over 2g^2}$
and $g^2$ is the continuum gauge coupling.
We will keep
$b^2={\beta\over N}$
fixed as $N$ goes to infinity,
and this amounts to setting
$g^2N={1\over 2b^2}$.

The reduced model on the lattice has a $[U(1)]^d$ symmetry.
The Eguchi-Kawai argument for reduction holds only if the
$[U(1)]^d$ symmetry remains unbroken~\cite{ek}.
This is not the case as $b^2\rightarrow\infty$~\cite{bhn}.
This problem is resolved by defining the quenched reduced
model~\cite{bhn} where
the eigenvalues of $U_\mu$,
\be
U_\mu = V_\mu D_\mu V^\dagger_\mu;\\
D_\mu = {\rm diag}(e^{i\theta^1_\mu}, e^{i\theta^2_\mu}, \cdots
e^{i\theta^N_\mu}),
\label{quenched}
\ee
are fixed, and $V_\mu$s are the only dynamical degrees of freedom.
The $\theta_\mu^i$ are randomly distributed in the interval $[-\pi,\pi]$,
and a quenched average over these variables is taken.
However in QCD$_2$, it is sufficient to
pick $\theta_\mu^i$ to be the $N$ roots of unity,
and there is no need to perform a quenched average~\cite{bhn}.
Minima of the action occur when all the $U_\mu$s are simultaneously
diagonal. Certain $V_\mu$s relate one minimum with another minimum.
Numerical simulations should sample all minima properly.

\begin{figure}
\epsfxsize = 0.45\textwidth
\centerline{\epsfbox{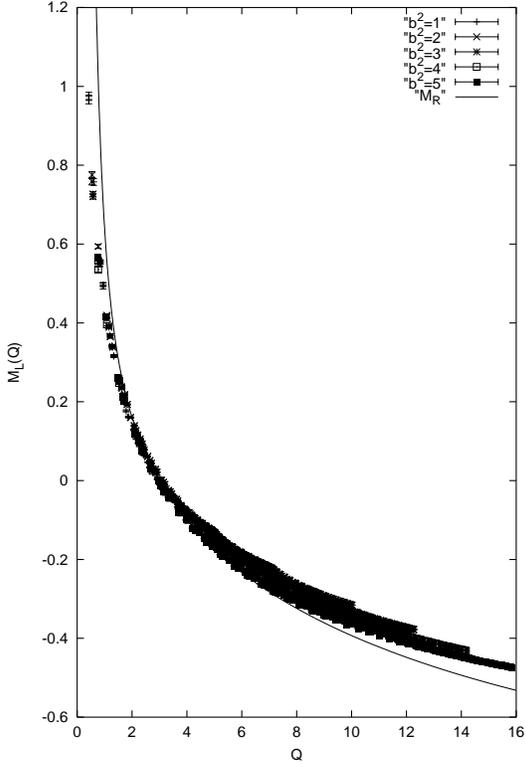}}
\caption{
Plot of the meson propagator for massless quarks
as a function of
$b^2$. The $b^2=1$ data is at $N_c=37$, the $b^2=2,4$ data are at $N_c=41$
and the $b^2=3,5$ data are at $N_c=47$.
}
\label{propg0}
\end{figure}

Fermionic loops are suppressed by $1/N$ in the large N limit, and only
the gauge degrees of freedom determine the dynamics. One can compute
the fermionic propagator in momentum space by force feeding momenta~\cite{ln}.
This
amounts to using $e^{ip_\mu}U_\mu$ for the fermion parallel transporter in
the $\mu$ direction which carries a momentum $p_\mu$.
We will use the propagator for massive overlap Dirac operator
given by~\cite{ehn}
\be
\label{massprop}
g(p)=\frac{1}{2m}
       \frac{1-\gamma_{d+1} \epsilon(H_w)}{(1+\mu) +(1-\mu) \gamma_{d+1}
\epsilon(H_w)}.
\ee
The quark mass is $m_q=2m\mu$ where $m$ is the magnitude of the
negative Wilson mass that appears
inside $H_w$, the hermitian Wilson-Dirac operator.

\section {Correlation functions}

\begin{figure}
\epsfxsize = 0.45\textwidth
\centerline{\epsfbox{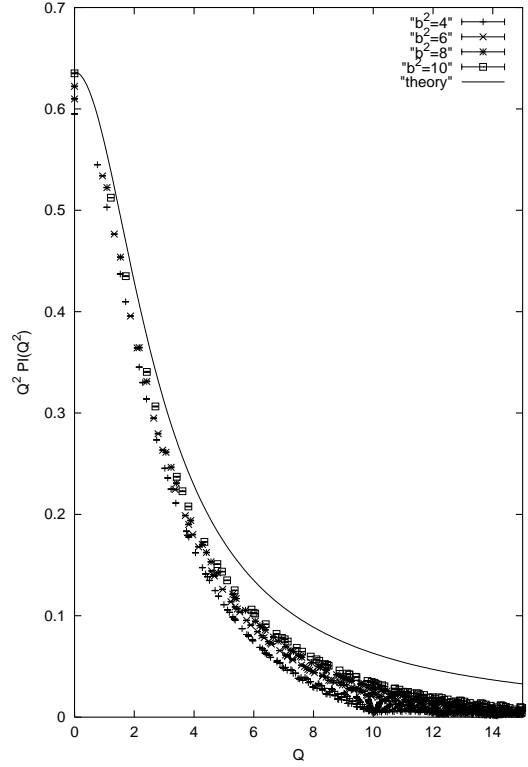}}
\caption{
Plot of the current-current correlator at $\gamma=1$
as a function of
$b^2$ at $N=41$.
}
\label{current}
\end{figure}

Meson correlation functions are known analytically in QCD$_2$~\cite{hooft},
and we can
check the results obtained from the quenched reduced model on the lattice.
Correlators can be expressed as an infinite sum over stable mesons, which
reproduce free field behavior at large momenta.
Therefore we have confinement and asymptotic freedom in QCD$_2$.
For $m_q=0$, there is a massless meson in the
spectrum, which is interpreted as a Goldstone boson.
Chiral symmetry breaking occurs if we take $m_q\rightarrow 0$
after we take $N\rightarrow\infty$.

Gauge fields are generated using (\ref{gauge}) and the
quenching condition in (\ref{quenched}).
The eigenvalues of $U_1$ and $U_2$ are set to the $N$ roots of unity,
and gauge field configurations are generated at a fixed $b^2$ for different
values of $N$. Fermionic momenta also are fixed at $N$ roots of unity.
The ``bare'' scalar and pseudoscalar propagators are computed using
\begin{eqnarray}
S_0 (P) &=& {1\over 4N^2} \sum_{p_1,p_2} \cr
&&Tr [ g(p_1+P_1,p_2+P_2) g(p_1,p_2)\cr
&+& g(p_1-P_1,p_2+P_2) g(p_1,p_2) \cr
&+& g(p_1+P_2,p_2+P_1) g(p_1,p_2) \cr
&+& g(p_1+P_2,p_2-P_1) g(p_1,p_2) ] \nonumber
\end{eqnarray}
\begin{eqnarray}
P_0 (P) &=& {1\over 4N^2} \sum_{p_1,p_2} \cr
&&Tr [ g(p_1+P_1,p_2+P_2) g^\dagger(p_1,p_2) \cr
&+& g(p_1-P_1,p_2+P_2) g^\dagger(p_1,p_2) \cr
&+& g(p_1+P_2,p_2+P_1) g^\dagger(p_1,p_2)\cr
&+& g(p_1+P_2,p_2-P_1) g^\dagger(p_1,p_2) ] \nonumber
\end{eqnarray}
This results in a correlation function for mesons that is a function
of one variable, namely, $P^2=4(\sin^2{P_1\over 2}+\sin^2{P_2\over 2})$.
The scale is set by $e^2={1\over 2\pi b^2}$, and we will use
$\gamma={m_q^2\over e^2}$ as a
dimensionless measure of the quark mass.
We will also use a dimensionless momentum $Q^2={P^2\over e^2}$.
The bare correlators have to  be regularized.  This is done by
a subtraction at zero momentum for massive quarks and a subtraction
at non-zero momentum for massless quarks. We will refer to the
renormalized quantities on the lattice by $S_L(Q^2,\gamma)$
and $P_L(Q^2,\gamma)$.
In the dimensionless notation,
we have to compare these results on the lattice with
the regularized results in the continuum
\be
P_R (Q^2, \gamma ) = \sum_{n\ge 0~{\rm even}}^\infty \left [
\frac{r_n^2}{Q^2 +\mu_n^2} - \frac {r_n^2}{\mu_n^2}\right ]
\ee
\be
S_R (Q^2, \gamma ) = \sum_{n\ge 1~{\rm odd}}^\infty \left [
\frac{r_n^2}{Q^2 +\mu_n^2} - \frac{r_n^2}{\mu_n^2} \right ] .
\ee
The dimensionless meson masses $\mu_n$ are
obtained by solving the 't Hooft Hamiltonian~\cite{paper},
and $\mu_n^2\sim \pi^2 n$ in the asymptotic limit.
The residues, $r_n^2$, are given by
\be
r_n^2 = {\gamma\over \pi} \Biggl[ \int_0^1 dx {\phi_n(x)\over x}\Biggr]^2
\ee
where $\phi_n(x)$ are the eigenvectors of the 't Hooft Hamiltonian.

Figure~\ref{pseudog1} shows the results for $P_L(Q^2,1)$ for various
values of $b^2$ and the comparison to the continuum result.
One can also compute the correlation function for massless quarks
at finite $N$. In this case, one will not see the effect of
chiral symmetry breaking. The propagators for the scalar and
the pseudoscalar are the same and should match with the average
of the two in the continuum. Figure~\ref{propg0} shows the
results for massless quarks along with the continuum result.
One can also compute  current-current correlators and obtain
$\Pi(Q^2)$. We present the results for current-current correlators
at $\gamma=1$ along with the continuum result in figure~\ref{current}.

\section {Conclusions}
The numerical results obtained in QCD$_2$ using the quenched reduced
model and overlap fermions are promising. The next step is to perform
the same computation for QCD$_4$, and this project is under way.

\end{document}